# Toughening β-$Ga_2O_3$ via mechanically seeded dislocations

Zanlin Cheng[1], Jiawen Zhang[1], Peng Gao[1], Guosong Zeng[1]*, Xufei Fang[2]*, Wenjun Lu[1]*

[1]*Department of Mechanical and Energy Engineering, Southern University of Science and Technology, Shenzhen, China*

[2]*Institute for Applied Materials, Karlsruhe Institute of Technology, Karlsruhe, Germany*

*Corresponding authors: zenggs@sustech.edu.cn (ZG); xufei.fang@kit.edu (XF); luwj@sustech.edu.cn (WL)

**Abstract:**

β-$Ga_2O_3$ is a promising candidate for next-generation semiconductors, but is limited by its intrinsic brittleness which hinders its application in flexible electronics and high-precision devices. This study explores a new approach to improving the damage tolerance of (001)-oriented β-$Ga_2O_3$ by introducing mechanically seeded dislocations via surface scratching. By applying a Brinell indenter to scratch the surface along the [100] direction, we effectively generate edge-type dislocations belonging to the (011)[01-1] and/or (0-11)[011] slip systems within a mesoscale a wear track. Through a combination of nanoindentation tests, surface morphology analysis, and microstructural characterization using scanning electron microscopy (SEM) and transmission electron microscopy (TEM), we reveal that the introduction of dislocations significantly mitigates the formation of cleavage cracks during indentation, in contrast to that observed in as-received β-$Ga_2O_3$. The mechanically seeded dislocations in the subsurface layers play an important role in preventing brittle fracture by facilitating stable plastic deformation.

## 1. Introduction

β-$Ga_2O_3$ is an emerging material for next-generation semiconductor applications, owing to its exceptional physical properties including an ultrawide bandgap of 4.6–4.9 eV[1], a high breakdown field strength of 8 MV/cm[2], and excellent thermal stability.[3] These characteristics make it an attractive candidate for use in high-power electronics, optoelectronic devices, and gas sensors.[4] Furthermore, large-size bulk β-$Ga_2O_3$ crystals can be grown using the economically viable melt-growth method, facilitating their potential for large-scale industrial applications.[5, 6]

However, despite its promising electrical properties, β-$Ga_2O_3$ suffers from inherent brittleness as is the case with most functional oxides, which restricts its mechanical reliability and performance, particularly in flexible electronic devices.[5] Like many other inorganic materials, β-$Ga_2O_3$ exhibits almost no room-temperature plastic deformation at the meso- or macroscale due to the intrinsically strong ionic and directional covalent bonds between constituent atoms, making it susceptible to cracking under mechanical loading.[7] The fracture occurs primarily along specific crystallographic planes, such as the (100) and (001) planes, which are known for their weakness in the crystal structure.[8] This challenge becomes more pressing during the fabrication of β-$Ga_2O_3$-based devices, where local fragmentation and cleavage fracture often occur during mechanical processes like grinding or polishing.[5]

To tackle the brittleness of functional oxides, various methods have been developed to improve the mechanical properties, particularly their fracture toughness and plasticity. These methods include surface treatment via dislocation engineering (mechanically seeded dislocations)[9-12] processing and fabrication,[13-15] and other techniques[9] aimed at introducing dislocations into crystals. Recent studies have shown that these mechanically seeded dislocations in brittle materials can effectively enhance their fracture toughness and plasticity at room temperature, especially in oxides.[16-19]

In contrast to the conventional view that dislocations are detrimental to semiconductors, engineering dislocations has been attracting new research attention for their capability of tuning versatile physical properties such as thermal conductivity,[20] photoconductivity,[21, 22] as well as superconductivity.[23] Another major advantage of introducing dislocations into the material through controlled mechanical processes is that it helps circumvent dislocation nucleation and facilitates dislocation multiplication and motion, thereby achieving large plasticity at room temperature.[16] However, this concept of mechanically seeded dislocations has so far only been validated on materials that were found to exhibit good room-temperature dislocation plasticity such as MgO, $SrTiO_3$, and $KNbO_3$,[16-19, 24-26] while its general applicability for truly brittle functional oxides remains unexplored at room temperature.

Here, we focus on β-$Ga_2O_3$ owing to its great potential for functional applications. β-$Ga_2O_3$ has a monoclinic structure (space group C2/m), consisting of tetrahedral $GaO_4$ units and octahedral $GaO_6$ units. There is a set of close-packed planes, (-201), (101), (310), and (3-10), leading to multiple candidates for the slip planes and Burgers vectors under concentrated mechanical loading.[27] Meanwhile, the existence of (100) and (001) cleavage planes makes them prone to cracking. The pertinent question arises: can the room-temperature mechanical properties of β-$Ga_2O_3$ be enhanced by using the simple approach of mechanically seeded dislocations?

To this end, here we investigate the impact of dislocations on the fracture toughness and room-temperature plasticity of (001) single-crystal β-$Ga_2O_3$. This particular crystal orientation is of importance as it has been widely adopted in the fabrication of Schottky barrier diodes (SBDs), a critical component in power electronic.[28, 29] We adopt a surface cyclic scratching technique along the [100] direction, which is perpendicular to the (100) and (001) cleavage planes, to introduce a network of dislocations beneath the surface. Nanoindentation tests were used to evaluate the effects of pre-engineered subsurface structure on mechanical properties. The morphology of imprints and cracks after nanoindentation was examined using scanning electron microscopy (SEM). Transmission electron microscopy (TEM) was used to analyze the dislocations generated by scratching. The post-mortem analysis beneath nanoindentation imprints using TEM underlines the role of pre-seeded dislocations on effective strengthening and crack-suppression mechanisms in β-$Ga_2O_3$.

## 2. Results and analyses

### 2.1 Microstructure modification by scratching

**Figure 1a** displays an overview surface morphology of the reference, as-received (001)-oriented β-$Ga_2O_3$ substrate. It indicates that the as-received (001)-oriented β-$Ga_2O_3$ substrates have a smooth surface without visible defects at macro/meso-scale. Detailed surface topographical analysis using scanning probe microscopy (SPM) with nanoscale spatial resolution reveals a mean surface roughness of 0.1 nm (**Figure 1b**). The cross-sectional annular bright-field (ABF) image in **Figure 1c** shows that no defects such as dislocations or stacking faults (SFs) exist in the β-$Ga_2O_3$ matrix. The selected-area diffraction pattern generated by the matrix corresponds well with the simulated standard diffraction pattern along the [010] zone axis, indicating a perfect crystal structure without lattice distortion, SFs, or impurities. However, there are two damaged layers above the β-$Ga_2O_3$ matrix (**Figure 1c**). The topmost layer, with a thickness of ~20 nm, is an amorphous phase, which can be confirmed by the absence of lattice fringes in high-resolution TEM (HRTEM) image (**Figure 1e**) and the weak diffraction ring in the Fast Fourier Transform (FFT) pattern (**Figure 1e$_1$**). Beneath the amorphous phase, the layer with a thickness of ~20 nm is composed of polycrystalline nanocrystals, as demonstrated by the locally ordered lattice fringes in the HRTEM image (**Figure 1f)** and the multiple diffraction spots from these nanocrystals in the FFT pattern (**Figure 1f₁)**. These damaged layers are likely introduced by chemical mechanical polishing (CMP) during the manufacturing process. Similar amorphous phases and nanocrystal layers induced by grinding and polishing have also been reported in β-$Ga_2O_3$.[30]

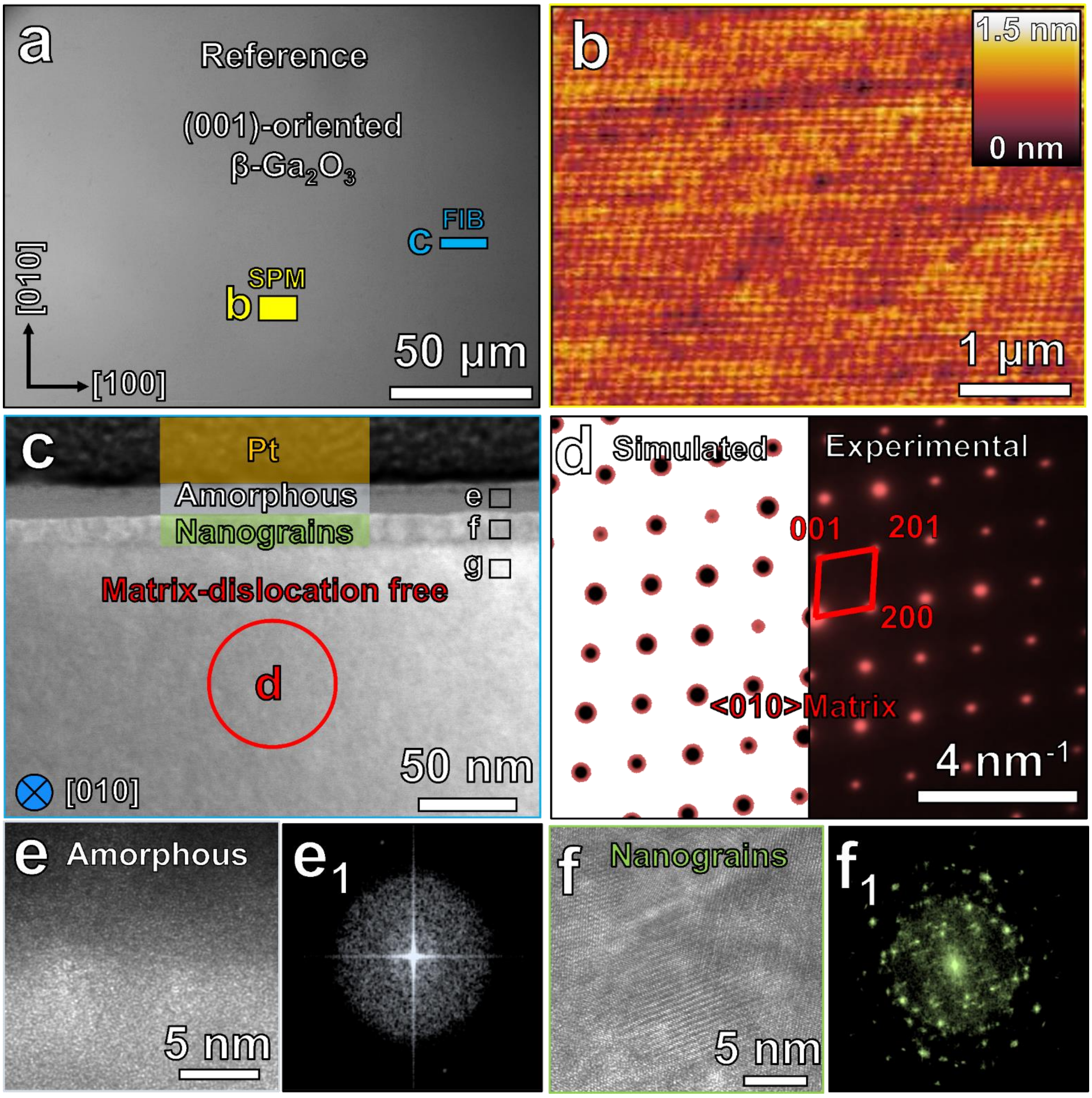

**Figure 1.** Surface morphology and microstructure of as-received β-$Ga_2O_3$ without scratching: (a) Representative optical microscopy image of the surface; (b) SPM topography image of the surface marked in (a); (c) Cross-sectional ABF image extracted from the region marked in (a); (d) Selected area diffraction pattern generated from the red circle in (b) and the corresponding simulated diffraction pattern; (e) and ($e_1$) HRTEM and FFT images of the amorphous phase layer marked in (b); (f) and ($f_1$) HRTEM and FFT images of the nanograin layer marked in (b).

Following the scratching process along the [100] direction under a load of 5 N, which involved 2000 reciprocal sliding cycles, a distinct wear track was introduced (**Figure 2a**). This trace was characterized by parallel furrows oriented along the [100] direction, reflecting the abrasive nature of the scratching process. According to further analysis of the wear track, obtained via white-light interferometry, no cleavage facets or brittle fracture were observed within the scratched trace (**Figure 2b**), suggesting that plastic deformation was the dominant mechanism during the scratching process. The depth profile of the wear track shows a maximum depth of ~80 nm over a width of 0.65 mm (**Figure 2c**), indicating a nominally flat surface. The lack of pile-up shoulders, a sign of plastic plowing, on both sides of the wear track indicates the limited extent of plastic deformation in $Ga_2O_3$ under the applied scratching conditions.

Surface topographical analysis performed by SPM revealed significant differences between the scratched and reference surface regions. The scratched surface exhibited an increase in surface roughness. After scratching, the mean roughness and root mean square (RMS) values were 0.8 and 1.2 nm, respectively (**Figure 2d**). This increase in roughness is a direct consequence of the interplay between abrasive debris and the sample. After scratching, a defect-rich subsurface region is created within the wear track. The cross-sectional ABF image in **Figure 2e** demonstrates that dislocations parallel to (001) plane and small dislocation loops with dark contrast (the initial of parallel dislocations) have been induced by scratching. In the magnified image shown in **Figure 2f**, similar to the top layers of as-received sample, amorphous phase and nanocrystalline region can be observed beneath the wear track. The top amorphous layer shows no lattice fringes in HRTEM image (**Figure 2g**) and weak diffraction circle in FFT pattern (**Figure 2g$_1$**). The second nanocrystal layer shows locally ordered lattice fringes in HRTEM image (**Figure 2h**) and the multiple diffraction spots from nanocrystals in the FFT pattern (**Figure 2h₁**). The thickness of these two layers is both ~20 nm, analogous to that in the reference, as-received sample, suggesting they are also induced by manufacture process instead of scratching.

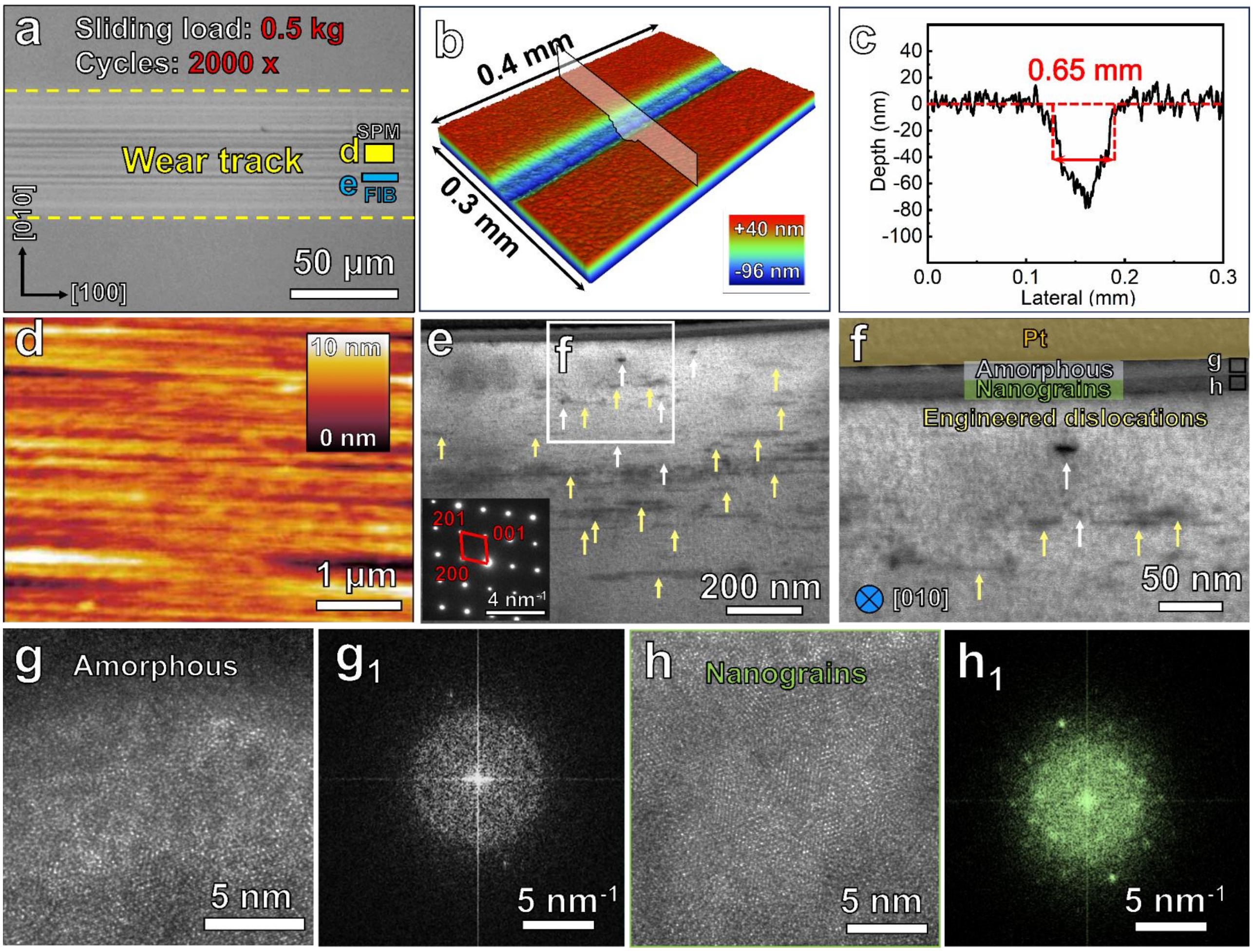

**Figure 2.** Surface morphology and microstructure of (001)-oriented β-$Ga_2O_3$ after scratching: (a) Representative optical microscopy image of the wear track; (b) 3D surface topography image of the wear track measured by white light interferometry; (c) 2D cross-sectional depth profile corresponding to the position marked by the inserted plane in (b); (d) SPM topography image of the wear track in the region marked in (a); (e) Cross-sectional ABF image extracted from the region marked in (a) (the insert shows the corresponding selected area diffraction pattern); (f) Magnified ABF image of the region marked in (e); (g) and ($g_1$) HRTEM and FFT images of the amorphous phase layer marked in (f); (h) and ($h_1$) HRTEM and FFT images of the nanograin layer marked in (f).

## 2.2 Hardness increase and crack suppression after scratching

**Figure 3** displays representative load-displacement curves (*P-h* curves) from nanoindentation tests as well as the morphology of nanoindentation imprints in the reference and scratched regions, with maximum loads ranging from 5 mN to 10 mN. Under the same load, the displacement in the wear track is lower than that in the reference region (**Figure 3a,b**), suggesting that the wear track has a higher hardness value. The inset I in **Figure 3a,b** displays the details of the curves at loads lower than 0.25 mN. The curves of the as-received sample overlap, while the curves obtained on wear track display scattering. It indicates that the as-received sample has a uniform crystal structure and surface quality while the surface quality as well as the subsurface structure beneath the wear track are not uniform. All nanoindentation curves of the as-received sample exhibit clear first pop-in events, characterized by a sudden displacement burst at a load of ~0.1 mN (**Figure 3a**). The first pop-in events occurred at displacement of 5-10 nm, signifying the transition from purely elastic to elasto-plastic deformation. This is further confirmed to correspond to dislocation nucleation in oxides,[31] since no evidence of phase transition was observed and the sharp Berkovich tip (with an effective tip radius of ~200 nm) used here suppresses crack formation.[31] The onset load of the first pop-in can then be used to calculate the maximum shear stress required for dislocation nucleation, as detailed in the **Supporting Materials-Section A**. The maximum shear stress of the as-received sample was calculated to be ~6.9 GPa (**Table 1**), lower than the upper bound of the theoretical shear strength of ~10.3 GPa required for homogeneous dislocation nucleation. The reason might be that the theoretical shear strength in specific slip systems has lower values because of the anisotropic mechanical properties of β-$Ga_2O_3$.[32, 33] Subsequent pop-in events appeared in the curves of the as-received sample when the load increased above 5 mN, as illustrated in region II (**Figure 3a**). Similar phenomena have been reported in the literature[34, 35], where continuous pop-in events were present in *P-h* curves during nanoindentation. These subsequent pop-in events are commonly related to the activation of multiple slip systems or crack initiation. The evidence can be found in the microstructure analysis in the following sections.

In contrast to the as-received sample, the *P-h* curves from the wear track exhibit no pop-in events. Similar phenomena were reported by Zhang et al. who found that *P-h* curves showed no pop-in events at low indentation depths after dislocations were induced in $SrTiO_3$ single crystals by scratching.[17] This suggests that the materials with mechanically seeded dislocations via cyclic scratching underwent more stable plastic deformation without the abrupt initiation of dislocation movement or crack formation. For β-$Ga_2O_3$, in addition to dislocations, plastic mechanisms including stacking faults (SFs) and crystal twins can also be activated due to their lower formation energy of 10-30 mJ/m$^2$ (for (100) SFs and (100) TBs observed in this work).[36] This will be further verified by microstructural characterization in **Sections 2.3** and **2.4**.

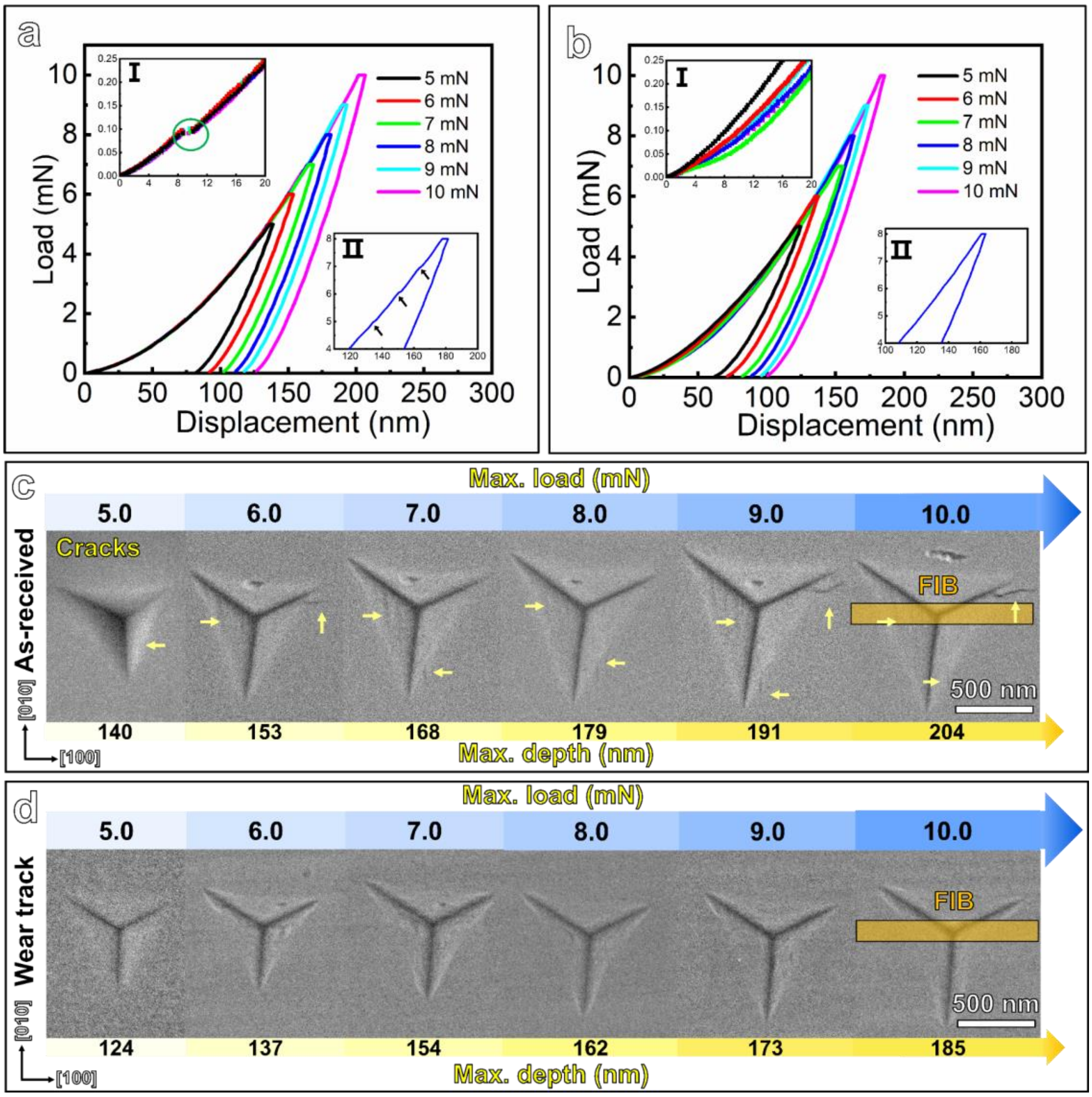

**Figure 3.** Load-displacement curves measured by nanoindentation on (a) as-received sample and (b) wear track with a max. load from 5mN to 10 mN; The insert I and II in (a) display the first pop-in (in green circle) and the subsequent pop-in events (pointed by black arrows) when the load is over 4 mN, while no pop-in events present in the insert I and II in (b); (c) and (d) Typical nanoindentation imprints on as-received surface and on wear track. The yellow arrows represent the cracks induced by nanoindentation.

The topography images of the nanoindentation imprints with a triangular shape (**Figure 3c,d**) show the residual impressions of the Berkovich diamond tip. Imprints on both the as-received surface and the wear track show an increase in size as a function of the indentation load from 5 mN to 10 mN. On the as-received surface, microcracks along [010] and [100] appeared after nanoindentation, in line with the previous reports on the easy cleavage planes in this material.[27, 35, 37] With increasing maximum load, the cracks became more pronounced and their length increased. In contrast, within the wear track, no cracks were observed in the nanoindentation imprints, which is direct evidence of effective suppression of brittle fracture or cleavage, as in the case of the as-received sample. Compared to the as-received sample, the maximum indentation depth on the wear track was tens of nanometers smaller, with correspondingly smaller indentation imprints. These features are direct evidence of the higher nanoindentation hardness after scratching.

The nanoindentation hardness (*H*) and elastic modulus ($E_s$) in **Table 1** for the as-received sample and wear track were obtained from *P-h* curves using Oliver-Pharr method.[38] Compared to the *H* value of 12.9 ± 0.1 GPa for the as-received sample, the value in the scratched region increased to 16.5 ± 1.0 GPa, while $E_s$ slightly increased from 204 ± 1 GPa to 219 ± 5 GPa after scratching. In what follows, we present evidence based on TEM analysis to explain the increased hardness caused by subsurface dislocations and microstructures induced by scratching, while the increase in elastic modulus of the scratched sample is likely caused by near-surface modification.

**Table 1** Nano-mechanical properties of as-received and scratched (001)-oriented β-$Ga_2O_3$.

| Material | Condition | Surface roughness (nm) | Hardness (GPa) | Youngs' modulus (GPa) | Max. shear stress (GPa) at first pop-in |
|---|---|---|---|---|---|
| (001)-oriented $Ga_2O_3$ | As-received | 0.1 ± 0.2 | 12.9 ± 0.1 | 204 ± 1.0 | 6.9 ± 0.3 |
| | After scratching | 0.8 ± 1.2 | 16.5 ± 1.0 | 219 ± 5.0 | Not applicable |

## 2.3 Characterization of subsurface damage after nanoindentation

To reveal the microstructural evolution beneath the nanoindentation imprints, TEM analysis was performed for both the as-received sample and the wear track. As illustrated in **Figure 4a**, after nanoindentation with a maximum load of 10 mN on the as-received sample, multiple slip bands were observed beneath the imprint. These slip bands extended to a depth of ~730 nm. The rotation of the diffraction spots away from the center spot indicates that the crystal lattice was deformed after nanoindentation (**Figure 4b**). According to the stereographic projection analysis in **Figure 4c**, the multiple slip bands were identified to be lie along (100), (101), and (-201) slip planes. In the region close to the indent tip contact point, no obvious SFs but a distorted lattice was observed (**Figure 4d**). Geometric Phase Analysis (GPA) analysis based on the HRTEM image in **Figure 4d** shows an inhomogeneous strain distribution in the distorted region. On the left side of the indented region, the DF image in **Figure 4f** acquired from the continuous streaks between diffraction spots shows bright stripes along (200) planes, suggesting that SFs align with the (200) crystal plane, which is consistent with the previous observations.[39, 40] On the right side of the indent, the deformation was dominated by dislocation slip bands in the (-201) plane (**Figure 4a**). A cleavage microcrack along (200) plane was observed in **Figure 4g** beneath the periphery of the imprint where internal stress built up during nanoindentation.[33] The asymmetric deformation patterns beneath the nanoindentation contact point were attributed to the asymmetric crystal structure of β-$Ga_2O_3$ and the activation of multiple slip systems, in line with previous reports.[33, 41, 42]

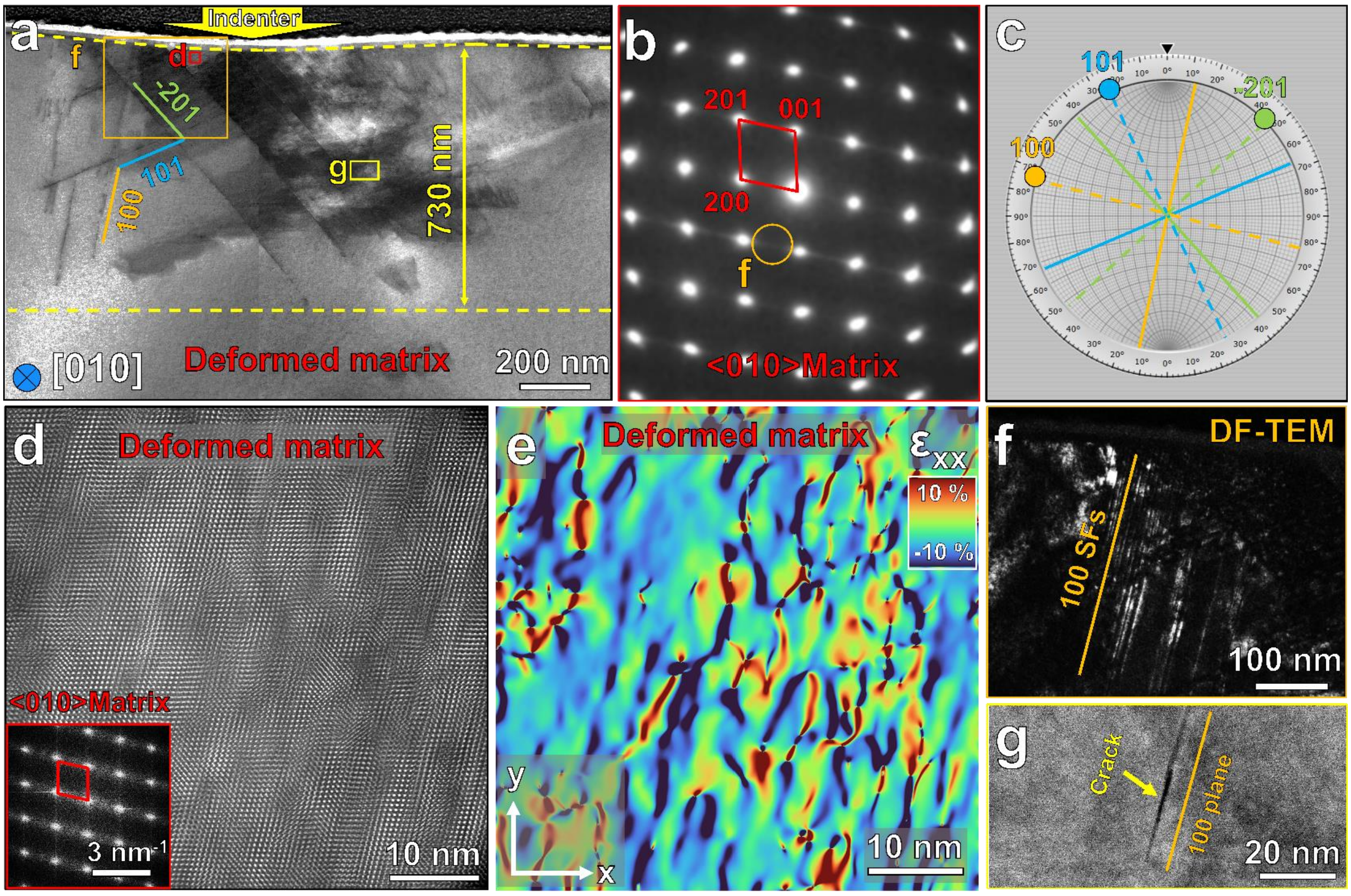

**Figure 4.** Cross-sectional TEM images of as-received sample after nanoindentation with a max. load of 10 mN: (a) ABF image showing the overview beneath the nanoindentation imprint; (b) Diffraction pattern of the matrix; (c) Stereographic projection analysis representing the slip traces and pole of planes; (d) HRTEM image of the region marked in (a); (e) GPA strain map corresponding to (d); (f) Dark-field image taken by the diffraction streaks marked by the circle in (b); (g) Magnified annular dark field (ADF) image of a crack along (100) plane in (a).

The microstructure after nanoindentation inside the wear track is illustrated in **Figure 5**. Similar to the as-received sample, after nanoindentation with a maximum load of 10 mN, slip bands along (100), (101), and (-201) were activated in the scratched sample (**Figure 5a**), which were indexed by the stereographic projection analysis in **Figure 5c**. The streaking diffraction spots along ***g*** of [100] (**Figure 5b**) indicate that the SFs along (200) plane were formed. The slip bands beneath the nanoindentation point also displayed asymmetric patterns (**Figure 5d,e**). However, the dislocation penetration depth of ~640 nm was lower than in the as-received sample, indicative of extensive dislocation interaction that contributes to the resistance against plastic deformation. On the right side of the indenter imprint, in contrast to the as-received sample in which microcracks existed and dislocation slip bands in the (-201) plane were the dominant feature, there is no microcrack beneath the imprint on the wear track and multiple slip bands are activated on the (101), and the (-201) and SFs along (200) plane in this region (**Figure 5e**). On the right side of the indenter tip, the SFs in DF image (**Figure 5f**) were more obvious and denser than those in the as-received sample (**Figure 4c**). Twins were found in the high-density SF region highlighted by the red solid square in **Figure 5g**, possessing a symmetric (100) plane, as corroborated by the mirror symmetry of lattice fringes and diffraction patterns in the FFT and HRTEM images (**Figure 5g**). For comparison, no twins were observed in the as-received sample. Multiple slip bands on the (101), and (-201) and SFs along the (200) plane interacted with each other and formed kinks (**Figure 5e,h**), inducing more severe inhomogeneous strain distribution (**Figure 5i**) than that in the as-received sample (**Figure 4e**).

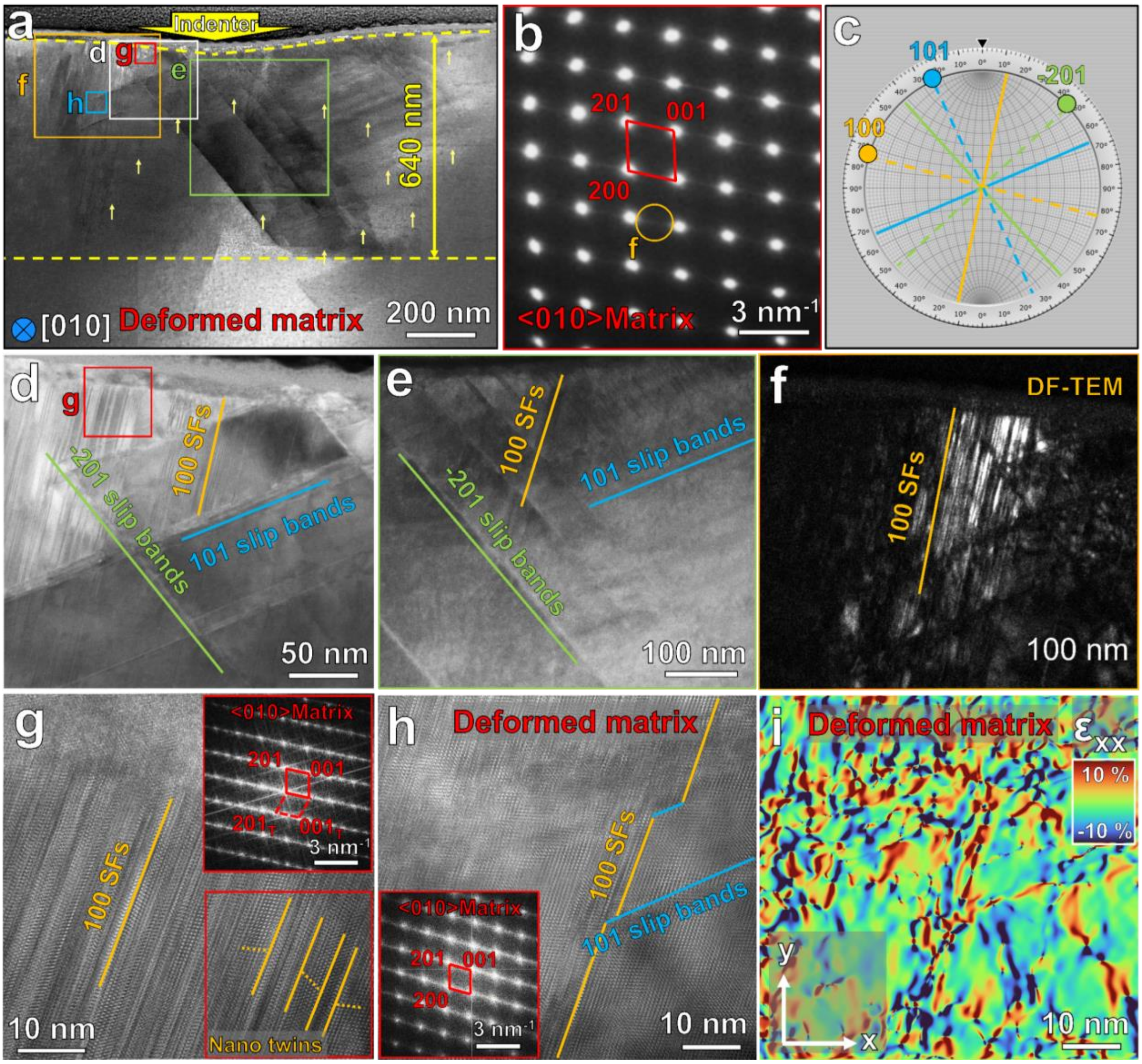

**Figure 5.** Cross-sectional TEM images of the nanoindentation imprint on wear track with a max. load of 10 mN: (a) ABF image showing the overview beneath the nanoindentation imprint on wear track (the yellow arrows mark the pre-engineered dislocations induced by scratching); (b) Diffraction pattern of the deformed matrix; (c) Stereographic projection analysis representing the slip traces and pole of slip planes; (d) and (e) ADF images demonstrating multiple slip systems in the deformed region marked in (a); (f) Dark-field image taken by the diffraction reflection marked by the circle in (b); (g) and (h) HRTEM images of the region marked in (a) (the insert shows the corresponding FFT image); (i) GPA strain map corresponding to (h).

## 3. Discussion

### 3.1 Dislocations induced by scratching

As demonstrated in **Section 2.1**, upon scratching (001)-oriented β-$Ga_2O_3$, dislocations parallel to (001) plane are mechanically generated beneath the scratched surface (**Figure 2**). To understand the role of the scratching-indued dislocations in crack suppression, it is essential to determine the dislocation types and slip system. Since β-$Ga_2O_3$ has a monoclinic crystal structure that belongs to space group C2/m, the low space symmetry of C2/m gives rise to a large number of possible Burgers vectors which are not crystallographically equivalent.[43] Its slip systems are more complex than other semiconductors that have high space symmetry. Hitherto, there are 7 slip planes and 11 possible Burgers vectors having been theoretically predicted and experimentally observed by using X-ray topography (XRT) and TEM in literature.[27, 30, 39, 41, 44-48] But most of them focused on the dislocations induced during crystal growth and post-growth cooling, in which the dislocation-generation mechanisms and dislocation types[49-52] are different from mechanically induced dislocations.

In order to identify the types and slip systems of these scratching induced dislocations, TEM-ADF images under two beam conditions were conducted. Under the seven ***g*** vectors displayed in **Figure 6a**, only one ***g*** vector made the parallel dislocations invisible, making it impossible to identify Burgers vectors using the conventional $\vec{g} \cdot \vec{b} = 0$ invisibility criterion, by which at least two inequivalent ***g*** vectors where the dislocation is out of contrast are required. However, the Burgers vector and slip plane can be determined by detailed analysis based on the and visibility conditions as listed in **Table S2**and the Schmid factors of possible slip systems in **Table 2**. As displayed in **Figure 6a**, the parallel dislocations were only invisible at ***g*** vector of [200] while visible at the others. According to the visibility criterion of dislocations with all possible Burgers vectors in **Table S2**, only three of them, which are ***b***=[001], [011] and [01-1], comply with the visibility under two beam conditions and the others inconsistent with the visibility in **Figure 6a** can be excluded.

The schematic illustration of stress condition during scratching is shown in **Figure 6b**. During cyclic scratching, the stress distribution beneath the spherical indenter is complex. In addition, some wear debris randomly falling on the wear track can create non-uniform and random stress distribution.[53]

To simplify the case for better understanding the relationship between the stress direction and the possible slip systems during scratching, the spherical indenter applied to the sample is equivalent to be normal force ($F_n$) and tangential force ($F_\tau$) on the affected region.[30] The normal force is perpendicular to the (001)-oriented surface with value of 5 N. Tangential force is the parallel to the surface and along the scratching direction [100] or [-100], given that reciprocating sliding motion was applied. The tangential force equals to the frictional stress (*f*) with a value of ~0.5 N, provided by the tribometer. Thus, the vector of the equivalent force can be determined. By calculating Schmid factors for the possible slip systems, the slip systems with high possibility to be activated can be identified.

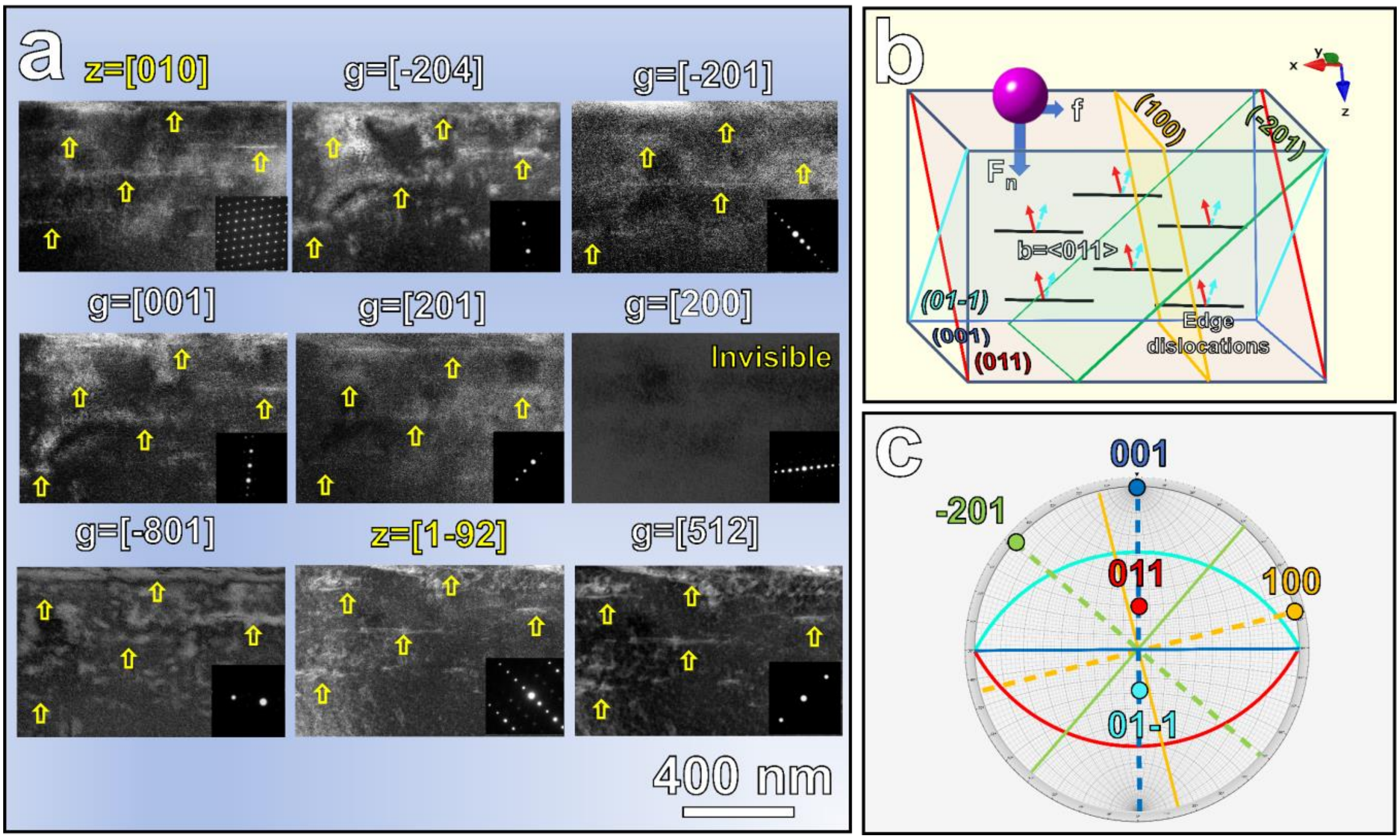

**Figure 6.** TEM analysis of dislocation types induced by scratching: (a) TEM-ADF images of dislocations under various zone axis (*z*=010 and *z*=1-92) and ***g*** vectors (***g***=-204, ***g***=-201, ***g***=001, ***g***=201, ***g***=200, ***g***=-801, and ***g***=512); (b) Schematic illustration of stress condition applied by spherical indenter during scratching, scratching-induced dislocations, and slip systems; (c) Stereographic projection analysis representing the slip traces and pole of slip planes.

The Schmid factors under the equivalent force during scratching and the corresponding slip systems are listed in **Table 2**. When the Schmid factor is (or close to) 0, the theoretical shear stress on the slip plane is (or close to) 0, which means that the angle between the external force and the slip direction or the normal of the slip plane is (or close to) 90°.[30] In this case, the shear stress caused by external force on the slip system is insufficient to provide enough driving force for dislocation glide in lattice to overcome the lattice resistance.[54] As displayed in **Table 2**, the slip systems with the Burgers vector of [010] or those on slip plane of (010), (-310), and (-3-10) have very low Schmid factors of (or close to) 0, indictive of the smaller likelihood of dislocations movement in these slip systems during scratching. It agrees with the TEM results that the Burgers vector of [010] can be safely excluded considering its visibility criterion under two beam conditions. No dislocation lines in TEM images align with the slip traces of (-310) and (-3-10) plane, no matter for the sample after scratching or after nanoindentation. In contrast, the slip systems in {011}<011>, {-201}<112>, (101)[10-1], and (001)[100] have relatively larger Schmid factors and some of them close to 0.5 --- the maximum value when the angle between the external force, the slip plane, and the direction was 45°. It indicates that the dislocations in these slip systems are more likely to glide during scratching. Given that the parallel dislocations in **Figure 6a** have line vector of [100] which is consistent with the slip traces of (011) and (0-11) slip planes as illustrated in **Figure 6c**, the parallel dislocations induced by scratching should be in (011)[01-1] and/or (0-11)[011] slip system. These dislocations should be edge type because the line vector of [100] is perpendicular to the Burgers vector. The spatial relationship between the dislocations and slip systems is demonstrated in the diagram in **Figure 6b**.

**Table 2** A summary of slip systems and dislocation types reported in literature[27, 30, 39, 41, 44-48] and the Schmid factors under the equivalent force during scratching (since the relationship between the applied force and some slip systems are not equivalent during the reciprocating motion, the Schmid factors with larger values are listed).

| Slip plane | Burgers vector | Dislocation-induced process | Schmid factors |
|---|---|---|---|
| (011) | [01-1] | Mechanical process: scratching on (001) plane | 0.41 |
| (01-1) | [011] | Mechanical process: scratching on (001) plane | 0.38 |
| (100) | [001] | Edge-defined film-fed growth (010)-oriented substrates;[44] Mechanical process: chemical mechanical polishing on (010)-oriented substrates;[30] Mechanical process: Nanopillar compression on (-201) -oriented single crystal[39] | 0.23 |
| | [010] | Edge-defined film-fed growth (010) and (-201)-oriented substrates;[44] | 0 |
| (-201) | [010] | Edge-defined film-fed growth of (001) [47, 48] and (-201)-oriented substrates[41, 46, 47] | 0 |
| | 1/2[112] | Edge-defined film-fed growth of (001)[48] and (-201)-oriented substrates[41, 46] | 0.48 |
| | 1/2 [1-12] | Edge-defined film-fed growth of (-201)-oriented substrates[41, 46] | 0.48 |
| (101) | [010] | Edge-defined film-fed growth of (-201)-oriented substrates[41, 46] | 0 |
| | [10-1] | | 0.49 |
| (001) | [100] | Edge-defined film-fed growth (010),[44] (001)[47, 48], and (-201)-oriented substrates[47] | 0.37 |
| | [010] | Edge-defined film-fed growth;[46] Mechanical process: Vickers indentation on the (001) surfaces[27] | 0 |
| (010) | [001] | Hydride-vapor-phase epitaxy growth (010)-oriented substrates[45] | 0 |
| (10-2) | [201] | Edge-defined film-fed growth (010)-oriented substrates[44] | 0.23 |
| (-310) | [001] | Edge-defined film-fed growth of (-201)-oriented substrates;[41, 46] Hydride-vapor-phase-epitaxy growth (010)-oriented substrates[45] | 0.14 |
| | 1/2 [130] | | 0.03 |
| | 1/2 [132] | | 0.04 |
| (-3-10) | [001] | Edge-defined film-fed growth of (-201)-oriented substrates[41, 46] | 0.14 |
| | 1/2 [130] | | 0.03 |
| | 1/2 [132] | | 0.04 |

### 3.2 Crack suppression and hardness increase

Nanoindentation testing was employed to assess the material's resistance to crack formation and its overall hardness after scratching. The as-received β-$Ga_2O_3$, which retained its pristine crystal structure, exhibited typical brittle fracture behavior. Upon application of a maximum load from 5-10 mN during nanoindentation, microcracks along the [010] and [100] directions were readily observed (**Figure 3**). These cracks formed due to the inherent brittleness of β-$Ga_2O_3$, which were particularly prone to cleavage fracture along (100) planes. In stark contrast, the nanoindentation imprints on the scratched β-$Ga_2O_3$ surfaces revealed a remarkable suppression of crack formation under maximum loads ranging from 5 mN to 10 mN (**Figure 3**), indicating that the scratching-induced dislocations can effectively inhibit the initiation of cracks. According to the microstructure beneath the imprints, cleavage microcracks along (200) plane can be found in the as-received sample (**Figure 4**), while there is no crack in the scratched sample with pre-seeded dislocations (**Figure 5**). Instead, higher density SFs along (100) plane, nanotwins, and slip bands in multiple slip systems are the dominate feature in the scratched sample (**Figure 5**). Thus, the crack suppression in the scratched sample should be attributed to the higher density (100)-oriented SFs, nanotwins, and multiple slip bands, all of which contribute to consuming the mechanical work during the nanoindentation tests. Similar phenomenon has been reported by Liu *et al.*[42] They performed nanoindentation on (100)-oriented β-$Ga_2O_3$ at both room temperature (RT) and high temperature (HT, 600 $^{o}$C). Higher density of SFs and nanotwins along (100) plane exist and the cracks were suppressed when the nanoindentation was conducted at HT, while obvious cleavage cracks, low density SFs and no nanotwins at RT.[42] These findings imply that the SFs and twins play a significant role next to dislocations in accommodating plastic deformation and allows for more ductile behavior instead of brittle failure in β-$Ga_2O_3$.

The formation of (100) twins in β-$Ga_2O_3$ induced by nanoindentation at ambient temperature remains quite rare reported.[33, 35, 55, 56] At elevated temperature, the presence of twins and high-density SFs along (100) plane can be attributed to the reduced barrier for the formation of twins and SFs. In the present work, it is reasonable to extrapolate that the scratching-induced dislocations plays a key role for their formation following heterogeneous nucleation mechanisms as for dislocations.[57, 58] Dislocations can act as sources for dislocation multiplication through Frank-Read sources and cross slip mechanisms.[57, 58] In materials with low SFs and TBs energy like β-$Ga_2O_3$, dislocations can act as potential nucleation sites for planar crystallographic defects.[59-65] Although up to now, the pre-existing dislocations induced planar defects and mechanisms are scarcely reported for β-$Ga_2O_3$, it has been widely investigated in other semiconductors such as 4H-SiC and its mechanism relates to dislocation dissociation,[61-65] but only for SFs. However, the promotion mechanisms for twins by pre-existing dislocation in semiconductors are still unclear although some evidence can be found in Al alloys, in which the substantial Shockley partial dislocations and stacking faults act as effective heterogeneous nucleation sites for nanotwins.[66, 67]

The toughening by mechanically seeded-dislocations in ceramics such as MgO, $SrTiO_3$, and $KNO_3$ has been reported recently.[17, 19, 24, 68] Contrast to MgO, $SrTiO_3$, and $KNO_3$, the effect of pre-seeded dislocations in β-$Ga_2O_3$ also relates to the promotion of SFs and twinning during nanoindentation owning to its low SF and twining energy (10-30 mJ/$m^2$).[36] The dislocations, SFs, and twins are effective in consuming the energy, release stress concentration, and suppress crack propagation[68] during the subsequence deformation such as nanoindentation tests. Meanwhile, the hardness values obtained from the load-displacement curves revealed that the scratched β-$Ga_2O_3$ exhibited significantly higher hardness compared to the as-received material. The hardness (*H*) of the scratched material was found to be 16.5 ± 0.97 GPa, compared to 12.9 ± 0.11 GPa for the as-received sample (**Table 1**).

The schematic illustration of crack suppression and hardening effects are displayed in **Figure 7**. In the as-received sample, no or few dislocations exist. When highly concentrated stress is applied by Berkovich indenter tip during nanoindentation, the sample is prone to fracture in a brittle manner even though low-density SFs and dislocations can be activated (**Figure 7a1, a2**). In contrast, the abundant pre-existing dislocations in the scratched samples can serve as sources for dislocation multiplication and motion. Moreover, high density SFs and twinning is easier to be activated (**Figure 7b1, b2**), although the synergistic effect between mechanically seeded dislocations, SFs and twins for the sustainable generation of more such defects remains unclear at this stage. Furthermore, the dislocations in the deeper layers of the scratched β-$Ga_2O_3$ contribute to hardening by impeding the movement of other dislocations. When dislocations encounter dislocations and other obstacles such as twins and SFs, they can be "pinned" to increase resistance for further plastic deformation (**Figure 7b2**), hence resulting in increased hardness as observed.[69] The higher dislocation density in the scratched material creates a more complex network of dislocations that resists the motion of dislocations under applied stress, thereby enhancing the material strength.[70, 71] However, the anisotropic mechanical properties and asymmetric crystalline structure emphasize the complexity between mechanical loading and microstructure evolution in nanoindentation experiments. A quantitative investigation is needed in future.

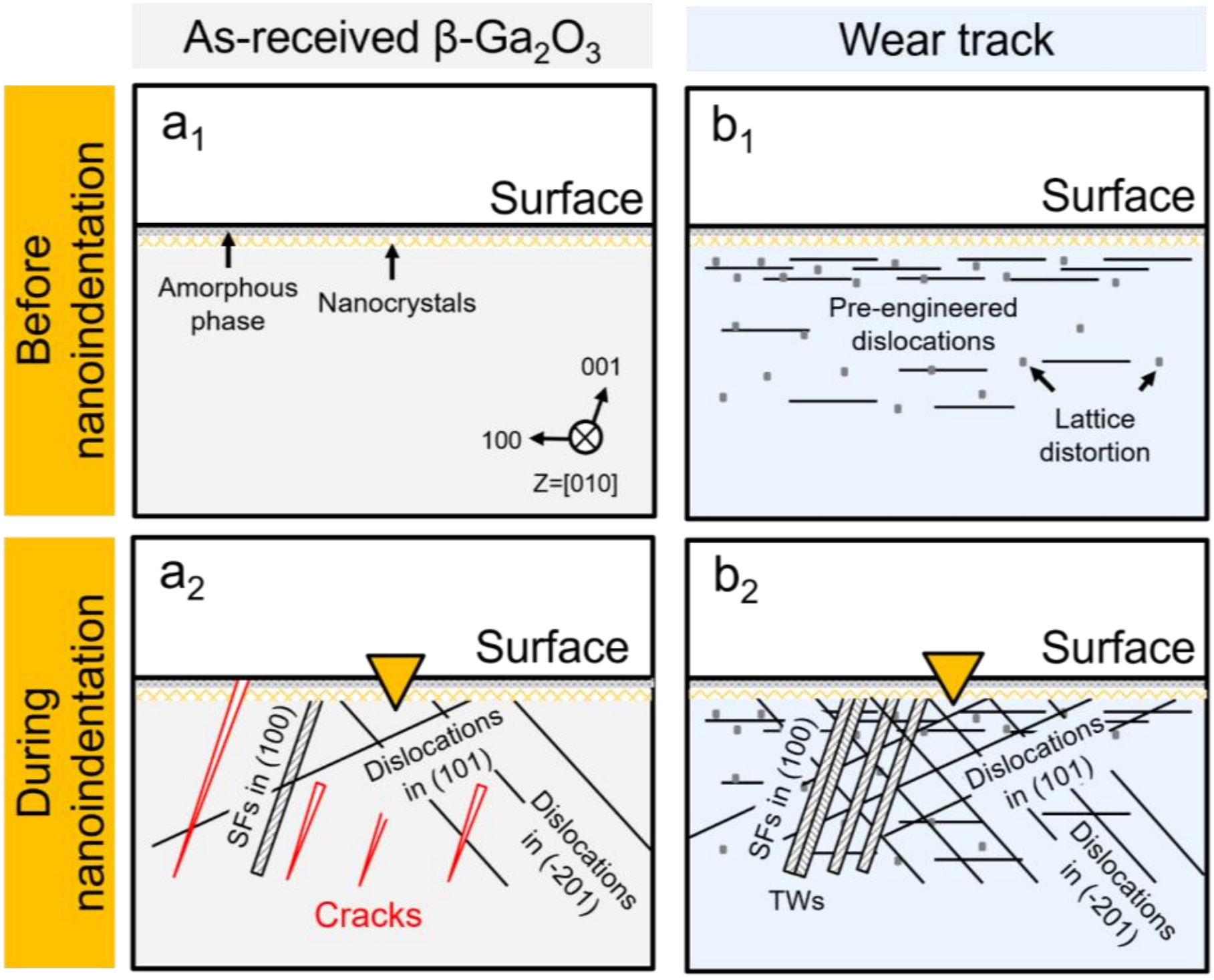

**Figure 7** Schematic illustration of nanoindentation on (a) as-received sample and (b) wear track with pre-seeded dislocations.

### 3.3 Mechanistic insights and implications for materials design

The mechanistic insight gained from this study provides a deeper understanding of the role that pre-engineered dislocations can play in enhancing the damage suppression of brittle functional oxides such as β-$Ga_2O_3$. The scratching-induced dislocations serve not only to increase the material's hardness but also alter its deformation mechanisms from brittle to ductile, although the length scale is still confined to micrometer. The mechanically induced pre-existing dislocations in the subsurface layers effectively facilitate the subsequent plastic deformation by suppressing cleavage fracture and promoting movements of plastic carriers including dislocations, SFs, and twins. This approach has significant implications for the design of next-generation semiconductor materials, particularly for applications in flexible electronics and high-power devices, where brittleness and crack formation are the key limitations. The findings of this study contribute to the broader field of material processing by providing a novel strategy for enhancing the damage tolerance and fracture toughness of intrinsically brittle oxides. The ability to engineer dislocations at the surface and subsurface regions offers a promising avenue for improving the machinability and reliability of materials used in various high-performance applications.

## 4. Conclusion

We demonstrate a novel approach to enhance the damage tolerance and mechanical properties of (001)-oriented β-$Ga_2O_3$ through the introduction of mechanically seeded dislocations via surface scratching with a Brinell indenter. Detailed TEM characterization reveals that the scratching-seeded dislocations are edge-type dislocations in the (011)[01-1] and/or (0-11)[011] slip systems. These dislocations effectively suppressed microcrack formation during subsequent nanoindentation testing. Simultaneously, nanoindentation also revealed that the scratched β-$Ga_2O_3$ exhibited much higher nanohardness compared to the as-received material. The absence of typical "pop-in" events in the load-displacement curves of the scratched samples further corroborates the stabilization of plastic deformation, facilitated by the seeded dislocations. These findings underline the critical role of dislocation engineering in mitigating brittle fracture and enhancing the overall mechanical performance of β-$Ga_2O_3$.

## 5. Experimental Section

*Material selection and sample preparation:* (001)-oriented β-$Ga_2O_3$ single crystals (provided by the 46$^{th}$ Research Institute of China Electronics Technology Group Corporation, China) were grown via the edge-defined film-fed growth method.[72] The samples have a dimension of 5 × 5 × 1 $mm^3$ and have been polished on one side, by which surface roughness lower than 1 nm is obtained and machining-induced defects such as dislocations and SFs have been successfully removed. The polished surface was subjected to surface scratching to induce a defined wear track. This was accomplished using a wear testing machine (Rtec, MFT2000, UK), equipped with a spherical $Al_2O_3$ ruby indenter (5 mm in diameter). The scratching procedure was performed under a normal load of 5 N to introduce mechanically seeded dislocations in near-surface region. A cyclic reciprocating sliding motion was applied, following the experimental protocol established in our previous studies.[10, 18] The experimental conditions were optimized to use 2000-cycle scratching to produce a wear track with dislocations engineered by plastic deformation without brittle fracture.

*Mechanical testing:* The mechanical properties of the β-$Ga_2O_3$ samples were characterized using nanoindentation[38, 73] (Hysitron TI Premier, BRUKER, USA). To minimize variables, nanoindentation tests with a diamond Berkovich indenter tip were conducted on the same sample in the undeformed region (reference) and on the wear track (with mechanically seeded dislocations) for direct comparison. The maximum loads vary from 5 mN to 10 mN, leading to a maximum depth of 110-210 nm. Loading and unloading time was 5 s, and the peak hold time was 2 s. For each indentation, the load-displacement data were collected and analyzed using the Oliver-Pharr method to extract the nanoindentation hardness and Young's modulus.[38] In order to ensure the accuracy and reproducibility of the results, over 20 repetitive indentation tests were performed for each test condition. To avoid the overlap of the plastic zones, the spacing of the indentation was set to be 10 μm, over 50 times of the indentation depth. By comparing the surface topographies of imprints as well as the mechanical properties of the reference region and on wear track, the effect of mechanically seeded dislocations on the material's resistance to crack propagation and nanohardness can be assessed. This approach is particularly relevant for materials subjected to micro-/nanoscale deformation, such as β-$Ga_2O_3$, where crack suppression is a critical factor in enhancing performance.

*Microstructure characterization:* To investigate the microstructural evolution of the treated β-$Ga_2O_3$ surfaces, a suite of characterization techniques was employed. The surface topography was first assessed using white light interferometry (Contour GT-K, Bruker, USA), providing high-resolution 3D images of the surface roughness and depth profile. Additionally, SPM equipped on nanoindentation was utilized to further explore surface morphology with a scan rate of 1 Hz and a contact force of 2 μN. The topographies of nanoindentation imprints on both the scratched and as-received samples was then analyzed using SEM (Apreo2 S Lovac) with secondary electron immersion mode at 5 keV and 0.1 nA. For detailed microstructural investigation, cross-sectional TEM foils were extracted at nanoindentation imprints in both reference and scratched regions using a dual-beam focused ion beam (FIB) equipped in an SEM (Helios Nanolab 600i, FEI, Hillsboro, USA). The TEM foils were milled under beam voltage of 30 keV and current of 0.79 nA, 0.43 nA, 0.23 nA, 80 pA, and 40 pA, sequentially. Then TEM images including ADF-STEM images, ABF-STEM images, selected area electron diffraction (SAED) patterns, and HRTEM were captured by using a TEM instrument (FEI Talos F200X G2, Thermo Fisher Scientific, USA) at an operating voltage of 200 keV. A probe semi-convergence angle of 10.5 mrad and inner and outer semi-collection angles of 23-55 mrad were used for ADF-STEM images, while 12-20 mrad were used for ABF-STEM images. The elaborated TEM characterization allows for direct observation of the defects induced during scratching and the subsequent nanoindentation, providing insight into dislocation densities, stacking faults, and crack formation mechanisms. FFT images based on HRTEM images are produced by using Velox software. The simulated diffraction pattern is generated by Single Crystal software. GPA was performed by using Strain++ software.

**Acknowledgements**

W. Lu acknowledges the support by Shenzhen Science and Technology Program (grant no. JCYJ20230807093416034), the Open Fund of the Microscopy Science and Technology-Songshan Lake Science City (grant no. 202401204), National Natural Science Foundation of China (grant no. 52371110) and Guangdong Basic and Applied Basic Research Foundation (grant no. 2023A1515011510). X. Fang thanks the support by the European Union (ERC Starting Grant, Project MECERDIS, grant No. 101076167). Views and opinions expressed are however those of the authors only and do not necessarily reflect those of the European Union or the European Research Council. Neither the European Union nor the granting authority can be held responsible for them. G. Zeng acknowledges the financial support from the Shenzhen Key Laboratory of Intelligent Robotics Flexible and Manufacturing Systems (No. ZDSYS20220527171403009) and the Shenzhen Science and Technology Innovation Commission (No.20231115111658002). The authors acknowledge using the facilities at the Southern University of Science and Technology Core Research Facility.

**Author contributions:**

**Zanlin Cheng:** Conceptualization, Formal analysis, Investigation, Methodology, Writing-original draft. **Jiawen Zhang:** Methodology, Data curation, Validation. **Peng Gao:** Methodology, Data curation, Validation. **Guosong Zeng:** Conceptualization, Methodology, Funding acquisition. **Xufei Fang:** Conceptualization, Methodology, Writing-review & editing, Supervision, Funding acquisition. **Wenjun Lu:** Conceptualization, Methodology, Writing-review & editing, Supervision, Funding acquisition.

**Conflict of Interest:** The authors declare no conflict of interest.

**Data Availability Statement:** The data that support the findings of this study are available from the corresponding author upon reasonable request.

**References:**

[1] T. Matsumoto, M. Aoki, A. Kinoshita, T. Aono, *Jpn. J. Appl. Phys.* **1974**, 13, 1578.

[2] K. Ghosh, U. Singisetti, *J. Appl. Phys.* **2018**, 124, 085707.

[3] X. Hou, Y. Zou, M. Ding, Y. Qin, Z. Zhang, X. Ma, P. Tan, S. Yu, X. Zhou, X. Zhao, G. Xu, H. Sun, S. Long, *J. Phys. D: Appl. Phys.* **2021**, 54, 043001.

[4] J. Yang, K. Liu, X. Chen, D. Shen, *Progress in Quantum Electronics*. **2022**, 83, 100397.

[5] S. J. Pearton, J. Yang, P. H. Cary, IV, F. Ren, J. Kim, M. J. Tadjer, M. A. Mastro, *Applied Physics Reviews*. **2018**, 5, 011301.

[6] M. Higashiwaki, *AAPPS Bulletin*. **2022**, 32, 3.

[7] Y. Q. Wu, S. Gao, R. K. Kang, H. Huang, *Journal of Materials Science*. **2019**, 54, 1958-1966.

[8] E. G. Víllora, K. Shimamura, Y. Yoshikawa, K. Aoki, N. Ichinose, *J. Cryst. Growth*. **2004**, 270, 420-426.

[9] X. F. Fang, *J. Am. Ceram. Soc.* **2024**, 107, 1425-1447.

[10] X. F. Fang, O. Preuss, P. Breckner, J. W. Zhang, W. J. Lu, *J. Am. Ceram. Soc.* **2023**, 106, 4540-4545.

[11] X. F. Fang, J. W. Zhang, A. Frisch, O. Preuss, C. Okafor, M. Setvin, W. J. Lu, *J. Am. Ceram. Soc.* **2024**, 107, 7054-7061.

[12] C. Okafor, K. Ding, O. Preuss, N. Khansur, W. Rheinheimer, X. F. Fang, *J. Am. Ceram. Soc.* **2024**, 1-14.

[13] P. Gao, R. Ishikawa, B. Feng, A. Kumamoto, N. Shibata, Y. Ikuhara, *Ultramicroscopy*. **2018**, 184, 217-224.

[14] J. Li, J. Cho, J. Ding, H. Charalambous, S. Xue, H. Wang, X. L. Phuah, J. Jian, X. Wang, C. Ophus, T. Tsakalakos, R. E. García, A. K. Mukherjee, N. Bernstein, C. S. Hellberg, H. Wang, X. Zhang, *Science Advances*. **2019**, 5, eaaw5519.

[15] K. Tsuji, Z. M. Fan, S. H. Bang, S. Dursun, S. Trolier-McKinstry, C. A. Randall, *J. Eur. Ceram. Soc.* **2022**, 42, 105-111.

[16] X. Fang;, W. Lu;, J. Zhang;, C. Minnert;, J. Hou;, S. Bruns;, U. Kunz;, A. Nakamura;, K. Durst;, J. Rödel, *Mater. Today*. **2024**, 82, 81-91.

[17] J. Zhang, O. Preuß, X. Fang, W. Lu, *JOM*. **2025**, 77, 3503-3512.

[18] J. Zhang, X. Fang, W. Lu, *Acta Mater.* **2025**, 291, 121004.

[19] O. Preuss, E. Bruder, W. J. Lu, F. P. Zhuo, C. Minnert, J. W. Zhang, J. Rödel, X. F. Fang, *J. Am. Ceram. Soc.* **2023**, 106, 4371-4381.

[20] J. Ding, J. Zhang, J. Dong, K. Higuchi, A. Nakamura, W. Lu, B. Sun, X. Fang, *Appl. Phys. Lett.* **2025**, 126, 253301.

[21] M. Soleimany, T. Frömling, J. Rödel, M. Alexe, *Adv. Funct. Mater.* **2025**, 35, 2417952.

[22] M. Kissel, L. Porz, T. Frömling, A. Nakamura, J. Rödel, M. Alexe, *Adv. Mater.* **2022**, 34, 2203032.

[23] S. Hameed, D. Pelc, Z. W. Anderson, A. Klein, R. J. Spieker, L. Yue, B. Das, J. Ramberger, M. Lukas, Y. Liu, M. J. Krogstad, R. Osborn, Y. Li, C. Leighton, R. M. Fernandes, M. Greven, *Nature Materials*. **2022**, 21, 54-61.

[24] O. Preuß, E. Bruder, J. Zhang, W. Lu, J. Rödel, X. Fang, *J. Eur. Ceram. Soc.* **2025**, 45, 116969.

[25] F. Appel, H. Bethge, U. Messerschmidt, *Physica Status Solidi (a)*. **1977**, 42, 61-71.

[26] N. Narita, K. Higashida, S. Kitano, *Scripta Metallurgica*. **1987**, 21, 1273-1278.

[27] H. Yamaguchi, S. Watanabe, Y. Yamaoka, K. Koshi, A. Kuramata, *Jpn. J. Appl. Phys.* **2022**, 61, 045506.

[28] K. Konishi, K. Goto, H. Murakami, Y. Kumagai, A. Kuramata, S. Yamakoshi, M. Higashiwaki, *Appl. Phys. Lett.* **2017**, 110, 103506.
[29] M. Higashiwaki, K. Konishi, K. Sasaki, K. Goto, K. Nomura, Q. T. Thieu, R. Togashi, H. Murakami, Y. Kumagai, B. Monemar, A. Koukitu, A. Kuramata, S. Yamakoshi, *Appl. Phys. Lett.* **2016**, 108, 133503.
[30] T. Hou, X. Ma, Y. Dong, P. Wang, Y. Li, Z. Jia, W. Mu, X. Tao, *Surfaces and Interfaces*. **2024**, 51, 104655.
[31] X. Fang, H. Bishara, K. Ding, H. Tsybenko, L. Porz, M. Höfling, E. Bruder, Y. Li, G. Dehm, K. Durst, *J. Am. Ceram. Soc.* **2021**, 104, 4728-4741.
[32] Q. An, G. Li, *Physical Review B*. **2017**, 96, 144113.
[33] Y. Z. Yao, Y. Sugawara, K. Sasaki, A. Kuramata, Y. Ishikawa, *J. Appl. Phys.* **2023**, 134, 215106.
[34] S. Gao, X. Yang, J. Cheng, X. Guo, R. Kang, *Mater. Charact.* **2023**, 200, 112920.
[35] T. Hou, W. Zhang, W. Mu, C. Li, X. Li, X. Ma, J. Zhang, H. Wang, Z. Jia, D. Liu, X. Tao, *Mater. Sci. Semicond. Process.* **2023**, 158, 107357.
[36] M. Wang, S. Mu, J. S. Speck, C. G. Van de Walle, *Advanced Materials Interfaces*. **2025**, 12, 2300318.
[37] X. Yang, Z. Dong, R. Kang, S. Gao, *Wear*. **2025**, 562-563, 205651.
[38] W. C. Oliver, G. M. Pharr, *J. Mater. Res.* **1992**, 7, 1564-1583.
[39] Y. Q. Wu, Q. J. Rao, J. P. Best, D. K. Mu, X. P. Xu, H. Huang, *Adv. Funct. Mater.* **2022**, 32, 2207960.
[40] Y. Q. Wu, S. Gao, H. Huang, *Mater. Sci. Semicond. Process.* **2017**, 71, 321-325.
[41] H. Yamaguchi, A. Kuramata, T. Masui, *Superlattices Microstruct.* **2016**, 99, 99-103.
[42] D. Liu, Y. C. Yan, Y. F. Bi, X. Gao, Q. Zhu, Y. Y. Liu, D. F. Wu, Z. Jin, N. Xia, H. Zhang, D. R. Yang, *J. Appl. Phys.* **2025**, 137, 125702.
[43] Gemma De la Flor Martin, G. Madariaga, *Acta Crystallographica Section A Foundations and Advances*. **2023**, 79, C326-C326.
[44] Y. Yao, Y. Ishikawa, Y. Sugawara, *Jpn. J. Appl. Phys.* **2020**, 59, 125501.
[45] M. Y. Kim, A. J. Winchester, A. F. Myers, E. J. Heilweil, O. Maimon, W. D. Yang, S. M. Koo, Q. Li, S. Pookpanratana, *Appl. Phys. Lett.* **2025**, 126, 231605.
[46] Y. Yao, Y. Sugawara, Y. Ishikawa, *J. Appl. Phys.* **2020**, 127, 205110.
[47] S. Masuya, K. Sasaki, A. Kuramata, S. Yamakoshi, O. Ueda, M. Kasu, *Jpn. J. Appl. Phys.* **2019**, 58, 055501.
[48] N. A. Mahadik, M. J. Tadjer, P. L. Bonanno, K. D. Hobart, R. E. Stahlbush, T. J. Anderson, A. Kuramata, *APL Materials*. **2019**, 7.
[49] J. Wittge, A. N. Danilewsky, D. Allen, P. McNally, Z. Li, T. Baumbach, E. Gorostegui-Colinas, J. Garagorri, M. R. Elizalde, D. Jacques, M. C. Fossati, D. K. Bowen, B. K. Tanner, *J. Appl. Crystallogr.* **2010**, 43, 1036-1039.
[50] B. K. Tanner, J. Wittge, D. Allen, M. C. Fossati, A. N. Danilwesky, P. McNally, J. Garagorri, M. R. Elizalde, D. Jacques, *J. Appl. Crystallogr.* **2011**, 44, 489-494.
[51] H. Wang, F. Wu, S. Byrappa, S. Sun, B. Raghothamachar, M. Dudley, E. K. Sanchez, D. Hansen, R. Drachev, S. G. Mueller, M. J. Loboda, *Appl. Phys. Lett.* **2012**, 100, 172105.
[52] B. Raghothamachar, M. Dudley, J. C. Rojo, K. Morgan, L. J. Schowalter, *J. Cryst. Growth*. **2003**, 250, 244-250.
[53] S. Li, P. Gao, G. Zeng, *Nanotechnology and Precision Engineering*. **2025**, 8, 033010.

[54] Z. W. Huang, P. L. Yong, H. Zhou, Y. S. Li, *Materials Science and Engineering: A*. **2020**, 773, 138721.
[55] R. Yang, N. Xia, K. Ma, D. Wu, J. Wang, Z. Jin, H. Zhang, D. Yang, *J. Alloys Compd.* **2024**, 978, 173556.
[56] S. Gao, X. Yang, X. Guo, J. Ren, R. Kang, *Mater. Charact.* **2023**, 206, 113441.
[57] D. J. Bacon;, D. Hull, *Introduction to Dislocations.* University of Liverpool.(UK: Elsevier. 2011), 48-50.
[58] J. J. Gilman, W. G. Johnston, *Dislocations in Lithium Fluoride Crystals.* Academic Press. 1962).
[59] I. Avci, M. E. Law, E. Kuryliw, A. F. Saavedra, K. S. Jones, *J. Appl. Phys.* **2004**, 95, 2452-2460.
[60] A. E. Romanov, P. Fini, J. S. Speck, *J. Appl. Phys.* **2003**, 93, 106-114.
[61] J. Nishio, C. Ota, R. Iijima, *Jpn. J. Appl. Phys.* **2022**, 61, SC1005.
[62] H. Tsuchida, I. Kamata, M. Nagano, *Journal of the Vacuum Society of Japan*. **2011**, 54, 353-361.
[63] J. J. Sumakeris, J. P. Bergman, M. K. Das, C. Hallin, B. A. Hull, E. Janzén, H. Lendenmann, M. J. O'Loughlin, M. J. Paisley, S. Ha, M. Skowronski, J. W. Palmour, C. H. Carter, *Mater. Sci. Forum*. **2006**, 527-529, 141-146.
[64] S. Ha, M. Benamara, M. Skowronski, H. Lendenmann, *Appl. Phys. Lett.* **2003**, 83, 4957-4959.
[65] H. Jacobson, J. Birch, R. Yakimova, M. Syväjärvi, J. P. Bergman, A. Ellison, T. Tuomi, E. Janzén, *J. Appl. Phys.* **2002**, 91, 6354-6360.
[66] V. Yamakov, D. Wolf, S. R. Phillpot, H. Gleiter, *Acta Mater.* **2002**, 50, 5005-5020.
[67] S. Liu, H. Ding, H. Zhang, R. Chen, J. Guo, H. Fu, *Nanoscale*. **2018**, 10, 11365-11374.
[68] M. N. Salem, K. Ding, J. Rödel, X. F. Fang, *J. Am. Ceram. Soc.* **2023**, 106, 1344-1355.
[69] J. Cho, Y. Li, Z. Shang, J. Li, Q. Li, H. Wang, Y. Wu, X. Zhang, *Materials Science and Engineering: A*. **2020**, 792, 139706.
[70] Y. Estrin, L. S. Tóth, A. Molinari, Y. Bréchet, *Acta Mater.* **1998**, 46, 5509-5522.
[71] A. Arsenlis, W. Cai, M. Tang, M. Rhee, T. Oppelstrup, G. Hommes, T. G. Pierce, V. V. Bulatov, *Modell. Simul. Mater. Sci. Eng.* **2007**, 15, 553.
[72] H. Aida, K. Nishiguchi, H. Takeda, N. Aota, K. Sunakawa, Y. Yaguchi, *Jpn. J. Appl. Phys.* **2008**, 47, 8506.
[73] X. Fang, A. Clausner, A. M. Hodge, M. Sebastiani, *MRS Bull.* **2025**, 50, 726-734.